\documentclass[%
      aps,
      prd,
      floatfix,
      preprintnumbers,
      preprint,
      showpacs,
      nofootinbib,
      tightenlines,
      amssymb,
      amsmath
]{revtex4-1}

\usepackage{url}
\usepackage{ifpdf}
\ifpdf
	\usepackage{cmap} % search in PDF
	\usepackage[pdftex]{color,graphicx}% include graphics
	\usepackage{epstopdf} %eps. -> .pdf figures
\else
	\usepackage{color,graphicx}
\fi
\usepackage{bm}% bold math
\usepackage[dvipsnames]{xcolor}
\usepackage{hyperref}
\hypersetup{%
colorlinks=true,%
citecolor=Blue,%
filecolor=Black,%
linkcolor=Blue,%
urlcolor=Violet
}
	
\usepackage{lineno}
\ifpdf
\usepackage[activate={true,nocompatibility},final,tracking=true,kerning=true,spacing=true,factor=1100,stretch=10,shrink=10]{microtype}
\fi
\begin{document}
\title{% \boldmath 
An Oxygen Target for (Anti)neutrinos 
}

\author{R.~Petti}
\email[]{Roberto.Petti@cern.ch}
\affiliation{Department of Physics and Astronomy, University of South Carolina, Columbia, South Carolina 29208, USA}

%%\date{\today}

\begin{abstract}

We discuss a method to obtain an effective oxygen target within a low-density detector allowing an accurate characterization of the 
various event topologies in $\nu (\bar \nu)$-oxygen interactions. Results can be of interest for long-baseline neutrino oscillation experiments 
utilizing water targets. In particular, the combination of both oxygen and hydrogen targets within the same detector can provide in-situ 
measurements of nuclear effects and of the (anti)neutrino flux, which are the leading sources of systematic uncertainties in long-baseline oscillation analyses.
These measurements can also provide useful information about the nuclear modifications of bound nucleons, as well as about the isospin symmetry in nucleons and nuclei.

\end{abstract}

%\begin{keyword} 
%Neutrino scattering, hydrogen target, straw tube tracker
%\end{keyword}

%\pacs{13.60.Hb, 12.38.Qk} %!Fix pacs!
\maketitle

\section{Introduction}
\label{sec:intro}

Measurements of high-energy neutrino interactions are challenging both at the source and at the detector sides. 
The high intensity of modern (anti)neutrino beams obviates the endemic lack of statistics of older neutrino experiments. However, the fact that the energy 
of the projectile (anti)neutrino is unknown on an event-by-event basis still represents an intrinsic limitation -- even when its overall energy spectrum is known with high precision -- 
making the detector itself the critical element in most cases. Besides factors like detector resolutions and energy scale uncertainties, the use of nuclei as (anti)neutrino 
targets appears ineluctably problematic. The initial momentum of the target nucleon within the nucleus is unknown and hadrons produced in the primary interaction 
can undergo an additional unknown modification as they can be absorbed or re-interact within the nucleus. Neutrino detectors have to infer the (anti)neutrino 
energy from the reconstructed final state particles emerging from the nucleus, which are affected by a substantial nuclear smearing and related systematic uncertainties. 

The issues above are exacerbated in long-baseline (LBL) neutrino oscillation experiments, in which the need of a multi-kton mass imposes heavy nuclear 
targets combined with relatively coarse detector resolutions. Their observation of CP violation in the leptonic sector relies on the detection of tiny differences between 
neutrino and antineutrino Charged Current (CC) interactions. Nuclear effects can introduce asymmetries between neutrinos and antineutrinos potentially mimicking the 
effect of CP violation, since they are in general isospin and flavor dependent. The physics sensitivity achievable by modern high-statistics LBL oscillation experiments 
is thus largely determined by their control of the various systematic uncertainties. In particular, the physics promise of next-generation projects like 
DUNE~\cite{Abi:2020evt} and Hyper-Kamiokande~\cite{Hyper-Kamiokande:2018ofw} is accompanied by an impressive percent-level precision required in systematics. 

Near detectors (ND) are the critical elements taking on the challenge of controlling systematic uncertainties in LBL experiments. To this end, they must fulfill two separate 
tasks characterized by conflicting requirements. On one side, they need the highest possible resolution -- together with a precise calibration of the energy scales and a broad acceptance -- in order to characterize in great details the various event topologies in $\nu (\bar \nu)$ interactions on the same nuclear target used in the far detectors (FD). 
The main goal is to provide in-situ measurements of nuclear effects and of the (anti)neutrino flux, which are typically the leading sources of systematic 
uncertainties in LBL oscillation analyses. On the other side, ideally they should also help to calibrate the event reconstruction in the FD, requiring an identical 
detector technology and necessarily a coarser resolution. This second task is complicated by the impossibility of having identical detectors at the near and far sites, 
due to differences in rates, event containment, (anti)neutrino energy spectra, etc. Such differences can imply sizable model corrections.  
As a result, reconstruction effects in the FD are often better controlled with dedicated test-beam exposures of the key detector elements, 
supplemented by appropriate in-situ calibration samples in the far detectors. Factorizing the two tasks using two separate ND technologies could help 
when only limited FD control samples are available. In the following we will focus on the primary ND task. 

Different nuclear targets have been used by LBL experiments, including lead (A=207) in OPERA~\footnote{The OPERA experiment was designed to detect 
$\nu_\tau$ appearance and did not have a near detector.}~\cite{Acquafredda:2009zz}, 
iron (A=56) in MINOS~\cite{MINOS:2008hdf}, carbon-based liquid scintillator ($<$A$>$=15.9) in NOvA~\cite{NOvA:2007rmc}, 
and water ($<$A$>$=14.3) in K2K~\cite{K2K:2006yov} and T2K~\cite{T2K:2011qtm}. 
Future projects will be based on an argon target (A=40) in DUNE and a water target in Hyper-Kamiokande, 
as well as in the THEIA proposal~\cite{Theia:2019non}. In principle, the use of light isoscalar nuclei like carbon or oxygen can benefit LBL measurements,  
even when the corresponding detector technologies have somewhat coarser resolutions. 
In all cases the near-detector measurements are a key factor in determining the ultimate physics sensitivity. 
In this paper we discuss a method to obtain an effective oxygen target based on a low-density detector allowing a precise characterization 
of nuclear effects and of the (anti)neutrino flux at the near-detector sites~\cite{Petti:2022bzt,Petti:2019asx,Duyang:2019prb,Petti:2023abz}. 

The paper is organized as follows. Sec.~\ref{sec:stt} briefly summarizes the detector technology designed to offer an accurate 
control of the neutrino targets. In Sec.~\ref{sec:oxigen} we discuss the ``solid" oxygen concept, while in Sec~\ref{sec:water} we 
describe different ways to obtain a corresponding water target. Section~\ref{sec:discussion} outlines the main features of those targets 
together with some examples of the physics measurements that they can enable.

\section{Control of Targets} 
\label{sec:stt} 

A detector technology designed to offer a control of the configuration, chemical composition, and mass of the neutrino targets similar to electron scattering 
experiments is a Straw Tube Tracker (STT), in which the targets are physically separated from the actual tracking system~\cite{Petti2004}. 
A large number (70-100) of thin planes -- each typically 1-2\% of radiation length $X_0$ -- of various passive materials with comparable thickness are alternated and 
dispersed throughout active layers -- made of four straw planes -- of negligible mass in order to guarantee the same acceptance 
to final state particles produced in (anti)neutrino interactions. The STT allows to minimize the thickness of individual active layers 
and to approximate the ideal case of a pure target detector -- the targets constitute about 97\% of the mass -- 
while keeping the total thickness of the stack comparable to one radiation length (Fig.~\ref{fig:STT_DesignConcept}). 
Each target plane can be removed or replaced with different materials during data taking, providing a flexible target configuration.  

The low average density $\rho \leq 0.17$ g/cm$^3$ and the overall dimensions comparable 
to one $X_0$ allow an accurate reconstruction of the four-momenta of the visible final state particles, as well as of the event kinematics in a plane transverse 
to the beam direction. The lightness of the tracking straws and the chemical purity of the targets, together with the physical spacing among the individual target planes, 
make the vertex resolution less critical in associating the interactions to the correct target material. For events with a single reconstructed charged track
the corresponding uncertainty is given by the ratio between the thickness of the straw walls ($<20 \mu m$) and the one of a single target layer, typically below 0.5\%.
For events with at least two reconstructed charged tracks this uncertainty is reduced to less than 0.1\%, 
thanks to a vertex resolution ($\ll 1$ mm~\cite{Anfreville:2001zi}) much smaller than the target thickness. 

\begin{figure}[tb]
\begin{center}
\includegraphics[width=1.00\textwidth]{STT_DesignConcept3}
\end{center}
\caption{Schematic drawing of STT modules (dimensions are in mm) equipped with the three replaceable targets~\cite{Petti:2022bzt,Petti:2019asx} listed in 
Tab.~\ref{tab:targets}:  polypropylene (left), polyoxymethylene (middle), and graphite (right). See text for details. 
}
\label{fig:STT_DesignConcept}
\end{figure}

The detector must be placed inside a magnetic field for the measurement of charged-particle momenta and surrounded by an electromagnetic calorimeter 
for the detection of photons, neutral hadrons including neutrons. The use of a distributed target mass within a relatively large volume ($\sim 40$ m$^3$) and a high track sampling 
of 0.15-0.30\% $X_0$ reduce the impact of multiple scattering on the measurements. The detector is optimized for the ``solid" hydrogen technique, in which 
$\nu(\bar \nu)$ interactions on free protons are obtained by subtracting measurements on dedicated graphite (C) targets from those on polypropylene (CH$_2$) 
targets~\cite{Petti:2022bzt,Petti:2019asx}. This technique is conceived to be model-independent, as the data from the graphite targets automatically include all 
types of processes, as well as detector effects, relevant for the selection of interactions on H.  For CC interactions the dilution factor with respect to a pure H$_2$ 
target can be reduced by a factor 5-7 with a kinematic analysis based on energy-momentum conservation~\cite{Duyang:2018xcc}. 
The thickness of the two default target materials, as well as the average density of the detector, depend on the value of the magnetic field available, in order 
to limit the multiple scattering contribution to the momentum and angular resolutions. For B=0.6 T we can use a thickness up to about 7 mm for the CH$_2$ 
targets and 4 mm for the C targets~\footnote{The C targets can be built from isostatic graphite, which is characterized by good mechanical properties, a density of about 1.8 g/cm$^3$, and a high purity.} (Fig.~\ref{fig:STT_DesignConcept}). 
Physics measurements are obtained effectively averaging over the many alternated targets, each of them individually 
transparent to final state particles (1-2\% $X_0$), making them relatively insensitive to construction tolerances. Every target element must be weighted 
before installation in order to control the total target mass with high accuracy. The detector operates at room temperature in a low-radiation environment, 
resulting in negligible aging effects on the thin passive targets. 

The STT performance was studied by simulating (anti)neutrino interactions with the GENIE~\cite{Andreopoulos:2009rq} and GEANT4~\cite{Agostinelli:2002hh} packages 
using the beam spectra expected at the future Long-Baseline Neutrino Facility (LBNF)~\cite{Abi:2020evt,Rout:2020cxi,LBNF-flux}. The active detector considered is composed 
of $\mathcal{O}(10^5)$ independent straws (Fig.~\ref{fig:STT_DesignConcept}) with maximal drift time of 60 ns. The average hit occupancy at LBNF is found to be less 
then 2\% including external backgrounds from interactions in the surrounding materials. We note that similar straws have been successfully used in much higher rate 
environments at the LHC~\cite{Abat:2008zz}. Simulations indicate that a single hit resolution of 200 $\mu m$ is sufficient for the various 
physics measurements. The average momentum resolution expected for muons is $\delta p / p \sim 3.5\%$ and the average angular resolution better than 2 mrad 
with the default CH$_2$ and C targets. The momentum scale can be calibrated to about 0.2\% using reconstructed $K_0 \to \pi^+\pi^-$ decays~\cite{Wu:2007ab,Duyang:2019prb}.
More details about the simulations and performance can be found in Refs.~\cite{Duyang:2018xcc,Duyang:2019prb,Petti:2022bzt}.

\section{Oxygen Target} 
\label{sec:oxigen} 

Since a pure oxygen target in liquid or gaseous form is not feasible due to safety and practical considerations, we are restricted to the oxygen available 
within chemical compounds. The precise control of the targets offered by the STT (Sec.~\ref{sec:stt}) allows the implementation of a ``solid" oxygen target 
from a subtraction between thin polyoxymethylene (CH$_2$O) and polypropylene (CH$_2$) targets. The former is an engineering 
thermoplastic (acetal, delrin) used for precision parts and characterized by high strength, hardness and rigidity, with $X_0 = 27.28$ cm and $\rho = 1.41$ g/cm$^3$. 
Several CH$_2$O planes can be easily integrated into the detector by replacing some of the default CH$_2$ targets. 
The distribution of the generic kinematic variables $\vec x$ in $\nu(\bar \nu)$-oxygen interactions can then be obtained as: 
\begin{equation} 
N_{\rm O}(\vec x) \equiv N_{\rm C H_2 O} (\vec x) - \frac{M_{\rm CH_2/CH_2O}}{M_{\rm C H_2}} N_{\rm C H_2 } (\vec x) 
\label{eq:Oevt} 
\end{equation} 
where $N_{\rm C H_2 O}$ and $N_{\rm C H_2}$ are the numbers of events selected from the polyoxymethylene and polypropylene targets, respectively. 
The interactions from this latter are normalized by the ratio between the total fiducial masses of CH$_2$ within the polypropylene and the acetal 
targets, $M_{\rm CH_2/CH_2O}/M_{\rm C H_2}$, which can be determined with high accuracy by weighting all relevant target elements with a precision scale. 
Both targets must have comparable thickness in terms of radiation and nuclear interaction lengths (Tab.~\ref{tab:targets}) 
and must be alternated throughout the detector volume to guarantee the same acceptance for final state particles. To this end, a solid acetal slab 4.5 mm thick can 
be used, corresponding to about 0.016 $X_0$. The oxygen content by mass within acetal is dominant at 53.3\%. 
Since polypropylene is the main target material required for the ``solid" hydrogen concept in STT, we expect the statistical uncertainty on the 
measured CH$_2$ background to be much smaller compared to the one of the acetal target. 
We note that the proposed technique could provide the first exclusive measurement of (anti)neutrino interactions on oxygen, as opposed to existing measurements which 
are performed on more complex chemical compounds like water.

\section{Water Targets} 
\label{sec:water} 

In addition to direct measurements on an oxygen target, having a complementary water target within the same detector can provide useful 
information for water-based detectors like Hyper-Kamiokande and THEIA (Sec.~\ref{sec:intro}). 
To this end, we can exploit the simultaneous presence of polyoxymethylene, polypropylene, and graphite targets in STT. The distribution of the generic 
kinematic variables $\vec x$ in $\nu(\bar \nu)$-water interactions can then be simply obtained from a subtraction between CH$_2$O and C targets: 
\begin{equation} 
N_{\rm H_2 O}(\vec x) \equiv N_{\rm C H_2 O} (\vec x) - \frac{M_{\rm C/CH_2O}}{M_{\rm C}} N_{\rm C } (\vec x) 
\label{eq:H2Oevt} 
\end{equation} 
where $N_{\rm C H_2 O}$ and $N_{\rm C}$ are the events selected from the polyoxymethylene and graphite targets, respectively. 
The interactions from this latter are normalized by the ratio between the total fiducial masses of C within the graphite and CH$_2$O targets, $M_{\rm C/CH_2O}/M_{\rm C}$. 
The advantages of this minimal approach are that we do not need to introduce additional targets, we can design all targets to have the same acceptance, 
and we avoid extraneous materials achieving a high chemical purity. The technique also provides water targets with a thickness $\leq$0.016$X_0$ (Tab.~\ref{tab:targets}), 
not achievable with standard liquid water. The water content by mass within acetal is 60\%. 
Similarly to the case of the oxygen target discussed above, the available mass of the graphite target is expected to be significantly larger than the 
C content within acetal, as it is an essential component of the ``solid" hydrogen technique. 
We note that the simultaneous presence of the three materials within STT would allow a complete characterization of the water target together 
with its separate constituent elements, O and H. 

We can also explicitly integrate thin water targets within STT, replacing some of the main polypropylene ones. Such passive water targets must be contained within 
sealed plastic shells. In order to minimize the total thickness of individual targets in terms of radiation length, as well as the amount of spurious materials to be subtracted 
from the shell, we can use 12 mm water layers encapsulated inside acetal shells 1.5 mm thick. The total effective thickness of such targets would be equivalent to 
about 0.044 $X_0$. The corresponding C content to be subtracted following Eq.(\ref{eq:H2Oevt}) to obtain a pure water target is only about 10.4\%. 
An interesting application of such water targets in STT is the measurement of $\nu$ and $\bar \nu$ interactions off the bound neutron in the deuteron (D), 
which can be obtained from a subtraction between heavy water (D$_2$O) and ordinary water (H$_2$O) targets~\cite{Petti:2022bzt}. 
To this end, both targets must be enclosed into identical acetal shells, which must be filled in such a way as to contain the same total mass of oxygen. 

\begin{table}[tb]
\begin{center} 
\begin{tabular}{|c|c|c|c|c|c|} \hline
 ~~~Target material~~~   &    ~Composition~  & ~Density~ & ~Thickness~ & ~Rad.\ length~  &  ~Nucl.\ int.\ length~  \\  
\hline\hline
Polypropylene & CH$_2$ & ~0.91 g/cm$^3$~ & 7.0 mm & 0.015 $X_0$ & 0.008 $\lambda_I$  \\ 
Graphite &  C  & 1.80 g/cm$^3$ &4.0 mm & 0.016 $X_0$  & 0.008 $\lambda_I$ \\ 
Polyoxymethylene & CH$_2$O  & 1.41 g/cm$^3$ & 4.5 mm & 0.016 $X_0$  &  0.008 $\lambda_I$ \\ 
\hline 
\end{tabular}
\caption{Possible parameters of the individual targets to be alternated within STT (for B=0.6 T) in the ``solid" oxygen and hydrogen techniques. 
The thickness can be fine-tuned depending on the specific detector configuration and application (Fig.~\ref{fig:STT_DesignConcept}). See text for details.} 
\label{tab:targets} 
\end{center} 
\end{table}

\section{Measuring Nuclear Effects} 
\label{sec:discussion} 

Nuclear effects and the (anti)neutrino flux are the leading sources of systematic uncertainties in high-energy neutrino scattering 
measurements~\cite{Petti:2019asx,NuSTEC:2017hzk}, as well as in long-baseline oscillation experiments~\cite{Coloma:2013tba,Mosel:2013fxa}. 
Both issues arise because in conventional (anti)neutrino beams the energy of the incoming neutrino is unknown on an event-by-event basis. 
The need to infer the neutrino energy from the detected final state particles constitutes an intrinsic limitation of high-energy neutrino experiments using nuclear targets, 
as the nuclear smearing introduces substantial systematic uncertainties in the process (Sec.~\ref{sec:intro}). The availability of both H and nuclear targets within 
the same detector can help to mitigate such problems in STT~\cite{Petti:2022bzt,Petti:2019asx}. The relative $\nu_\mu$ and $\bar \nu_\mu$ fluxes as a function 
of energy can be determined in-situ with an accuracy around 1\% using exclusive $\nu_\mu p \to \mu^- p \pi^+$ and $\bar \nu_\mu p \to \mu^- n$ processes 
on H at small energy transfer~\cite{Duyang:2019prb}. The combined use of $\nu$-H and $\bar \nu$-H CC interactions can provide a control sample free from 
nuclear effects to calibrate the neutrino energy scale in CC interactions from the nuclear targets~\cite{Petti:2022bzt}. 

The STT offers a tool to measure nuclear modifications of cross-sections and to constrain the systematic uncertainties associated to the nuclear 
smearing for the various integrated nuclear targets. Each individual target is designed to be transparent to final state particles (Tab.~\ref{tab:targets}) allowing, 
together with the low average density of the detector $\rho \leq 0.17$ g/cm$^3$ and the high track sampling 0.15-0.30\% $X_0$, 
an accurate reconstruction and characterization of the various event topologies in $\nu (\bar \nu)$ interactions. 
Simulations of the detector response with neutrino interactions using the LBNF beams (Sec.~\ref{sec:stt}) result in a rather uniform acceptance over 
the full 4$\pi$ angle, with values of 95-99\% for $\mu^\pm, \pi^\pm, K^\pm, e^\pm$. Charged tracks can be reconstructed down to momenta of about 50 MeV/c 
for $\mu^\pm, \pi^\pm$ and about 10 MeV/c for $e^\pm$. The reconstruction efficiency for protons and neutrons is mostly affected by the thickness of the 
individual passive targets. The values in Tab.~\ref{tab:targets} result in an average efficiency of about 66\% and 94\% for protons originated in inclusive 
$\nu_\mu$ CC interactions on carbon/oxygen and hydrogen targets, respectively~\footnote{In principle some of the protons 
which do not have enough hits for the momentum reconstruction could be still detected from the energy deposition in the crossed straws. Such cases are not considered here.}. 
Neutrons can be reconstructed from the detection of secondary particles produced from 
interactions in both the STT elements and in the surrounding electromagnetic calorimeter. The average neutron detection efficiencies are about 74\% and 82\% 
for $\bar \nu_\mu$ CC interactions on carbon/oxygen and hydrogen targets, respectively. 

A key requirement is to guarantee the same acceptance across all nuclear targets, which is achieved by the 
combined effect of their thinness (Tab.~\ref{tab:targets}) and of their alternation throughout the detector volume. Detailed detector simulations 
indicate that in this way the acceptance difference between targets can be kept within $10^{-3}$ for all particles. 
The subtraction procedure required to obtain interactions on H, O, and H$_2$O can then be considered model-independent. Furthermore, the detector 
acceptance effectively cancels out in comparisons among the selected interactions on the H, C, and O targets. 

\begin{table}[tb]
\begin{center} 
\begin{tabular}{|l|c|c|c|c|c|} \hline
    &  ~~Fiducial~~ &  \multicolumn{2}{|c|}{CP optimized}
    &     \multicolumn{2}{|c|}{$\nu_\tau$ optimized}   \\ 
 ~~Target & mass (t) & ~~$\nu_\mu$ CC FHC~~ &  ~~$\bar \nu_\mu$ CC RHC~~  &  ~~$\nu_\mu$ CC FHC~~ &  ~~$\bar \nu_\mu$ CC RHC~~  \\ \hline\hline
~~CH$_2$   & 3.60  & 13,589,000~  &  5,087,000   &  32,472,000~  &    11,650,000~    \\ 
~~C    & 0.60 & 2,397,000  &   ~~806,000   &  5,720,000  &  1,863,000   \\ 
~~CH$_2$O   & 1.40  &  5,449,000 &   1,927,000   &  13,012,000~  &   4,433,000  \\ 
\hline\hline 
~~O (subtracted) ~~   &  0.76  &  2,982,850  &  1,003,800    &   7,118,940 &  2,318,740   \\ 
                                          &    &  ~~$\pm$ 2,428 &   ~~$\pm$  1,447 &  ~~$\pm$ 3,752 &  ~~$\pm$  2,195 \\ 
~~H$_2$O (subtracted)~~ & 0.85 &  3,211,800 &   1,174,730   &  7,673,330  &  2,694,200   \\ 
                                          &    &  ~~$\pm$ 2,745 &   ~~$\pm$  1,621 &  ~~$\pm$ 4,242 &  ~~$\pm$  2,461 \\ 
\hline 
\end{tabular}
\caption{Number of inclusive CC events expected on the various targets considered with both the low-energy and high-energy beam spectra~\cite{LBNF-flux} available at the  
future LBNF. We assumed a fixed exposure of $3\times10^{21}$ protons on target  for both the neutrino (FHC) and antineutrino (RHC) beams, 
a detector located at 574 m from the source, and the replacement of about 30\% of the main CH$_2$ targets with equivalent CH$_2$O ones (Tab.~\ref{tab:targets}). 
The bottom rows provide the corresponding number of events on the resulting oxygen and water targets with the expected statistical uncertainties after subtraction. 
See text for details. 
} 
\label{tab:events} 
\end{center} 
\end{table} 

The high intensity of modern (anti)neutrino beams complements well the relatively small mass of the various targets in STT. 
For illustration, a fiducial mass of one tonne of water at the future LBNF will collect about $1.5\times 10^6$ $\nu_\mu$ CC events/year 
with the default low-energy spectrum~\cite{LBNF-flux} (a factor of two higher with the planned PIP-II upgrade) and about $7\times 10^6$ $\nu_\mu$ CC events/year 
with the high-energy beam spectrum~\cite{LBNF-flux} and the upgraded beam~\footnote{On-axis rates expected at the near-detector site located at 574 m from the source.}. 
With such high event rates a limited number of acetal and/or water targets in STT would suffice to 
obtain sensible physics measurements. Assuming as a reference a STT configuration with a ``solid" hydrogen mass equivalent to about 10 m$^3$ of liquid 
H$_2$~\footnote{A fiducial mass of ``solid" hydrogen around 700 kg can be obtained from the combination of about 5 tons of polypropylene and about 600 kg of graphite.}, 
it is relatively easy to achieve an overall water target mass close to one tonne.
As an example, replacing around 30\% of the main CH$_2$ targets (corresponding to about 20 STT modules) with the equivalent CH$_2$O ones described 
above (Tab.~\ref{tab:targets}) would provide concurrently an oxygen target mass of about 760 kg and a water target mass of about 850 kg. 
Table~\ref{tab:events} summarizes the expected number of $\nu_\mu$ CC and $\bar \nu_\mu$ CC events with such a detector configuration 
at the LBNF for a fixed exposure of $3\times 10^{21}$ protons on target, achievable in about two years with the default beam intensity and in about 
one year with the planned beam upgrades.  

\begin{figure}[tb]
\begin{center}
\includegraphics[width=0.50\textwidth]{nuke_Rnuah_F2O}\hspace*{0.10cm}\includegraphics[width=0.50\textwidth]{nuke_Rnuah_F3O}
\end{center}
\caption{Ratios $R_2^{\rm \nu O}$ (left panel) and $R_3^{\rm \nu O}$ (right panel) as a function of the Bjorken $x$ for different values of the momentum transfer 
$Q^2=1,2,5,10,20$ GeV$^2$. See text for details.
}
\label{fig:nuke-OH}
\end{figure}

\begin{figure}[tb]
\begin{center}
\includegraphics[width=0.50\textwidth]{isospin_F2Ac_Q2}\hspace*{0.10cm}\includegraphics[width=0.50\textwidth]{isospin_F3Ac_Q2}
\end{center}
\caption{Ratios ${\cal R}_2^{\rm A}$ (left plot) and ${\cal R}_3^{\rm A}$ (right plot) in the O and C isoscalar nuclei as a function of the momentum transfer $Q^2$. 
The various curves correspond (from bottom to top) to values of $x=0.05,0.1,0.15,0.2,0.25,0.3,0.35,0.4,0.5,0.6,0.7,0.8$. 
See text for details.
}
\label{fig:isospin-OC}
\end{figure}

The statistics from Tab.~\ref{tab:events} in combination with the 4$\pi$ acceptance and the high momentum and angular resolutions of the detector (Sec.~\ref{sec:stt}) 
allows precision measurements of exclusive and inclusive topologies in both $\nu_\mu$ CC and $\bar \nu_\mu$ CC interactions, as well as of rare processes 
characterized by tiny cross-sections. The subtraction procedure introduces a relatively modest increase of about 40\% in the statistical uncertainties with 
respect to the case of ideal targets (Tab.~\ref{tab:events}). The dominant systematic uncertainties are expected to be related to the knowledge of the energy scales and 
of the incoming (anti)neutrino fluxes. The momentum scale can be calibrated to about 0.2\% using reconstructed $K_0 \to \pi^+\pi^-$ decays~\cite{Wu:2007ab,Duyang:2019prb}, 
resulting in energy scale uncertainties roughly comparable with the statistical ones with the samples in Tab.~\ref{tab:events}. 
The (anti)neutrino fluxes as a function of energy can be determined with an accuracy around 1\% using exclusive $\nu_\mu H \to \mu^- p \pi^+$ and 
$\bar \nu_\mu H \to \mu^+ n$ processes~\cite{Duyang:2019prb,Petti:2023abz} and are expected to represent the leading source of systematics. 

Comparing measurements of the average bound nucleon structure functions $F_{2,3}^{\rm \nu O}$ from the ``solid" oxygen with the ones of the 
free nucleons in H with similar acceptance through the ratio $R_{2,3}^{\rm \nu O} = 2 F_{2,3}^{\rm \nu O} / \left( F_{2,3}^{\rm \nu H} + 
F_{2,3}^{\rm \bar \nu H} \right)$ can provide insights on the nuclear modifications of the nucleon properties~\cite{Petti:2022bzt,Kulagin:2004ie,Kulagin:2007ju,Kulagin:2014vsa}. 
The ratios $R_2^{\rm \nu O}$ and $R_3^{\rm \nu O}$ are shown in Fig.~\ref{fig:nuke-OH} as a function of the Bjorken $x$, 
calculated at the NNLO approximation in the strong coupling constant, using the results of the global QCD analysis of Ref.~\cite{Alekhin:2007fh}
and the nuclear model of Refs.~\cite{Kulagin:2004ie,Kulagin:2007ju,Kulagin:2014vsa}. 
The oxygen target can also provide complementary measurements with respect to the C and Ca targets to test the isospin (charge) symmetry~\cite{Petti:2022bzt}. 
The isotopic content expected for a standard O target is 99.76\% of ${}^{16}$O, 0.2\% of ${}^{18}$O, and 0.04\% of ${}^{17}$O, resulting on average in the smallest 
isovector component among stable elements $\beta = (2Z-A)/A = 6 \times 10^{-5}$. A comparison between $\nu$ and $\bar \nu$ interactions on oxygen through the ratios 
${\cal R}_2^{\rm O} = F_2^{\bar \nu} / F_2^{\nu} - 1$ and ${\cal R}_3^{\rm O} = xF_3^{\bar \nu} / xF_3^{\nu} - 1$ for the structure functions $F_2$ and $xF_3$ 
can provide useful information about the isospin symmetry in nucleons and nuclei. 
Figure~\ref{fig:isospin-OC} shows the values of such ratios for both oxygen and carbon as a function of the momentum transfer $Q^2$ for different values of the 
Bjorken $x$.

%\begin{acknowledgments}

%\end{acknowledgments} 

%\newpage
%%% REFERENCES
% you should have RevTeX bibliography styles in the search path
%\bibliographystyle{apsrev}
\bibliography{main1}

\end{document}